\pdfoutput=1
\documentclass[11pt,letterpaper]{article}
\usepackage{naaclhlt2016}
\usepackage{times}
\usepackage{microtype}
\usepackage{latexsym}
\usepackage{multirow}
\usepackage{multicol}
\usepackage{booktabs}
\usepackage{hyperref}

\usepackage{graphicx}
\usepackage{amsmath} 
\usepackage{algorithm2e}
\usepackage{algpseudocode}
\usepackage{color}
\usepackage{bold-extra}
\usepackage[T1]{fontenc}

\naaclfinalcopy 

\title{Improving Document Clustering by Eliminating Unnatural Language}

\author{Myungha Jang$^{1}$, Jinho D. Choi$^{2}$, James Allan$^{1}$\\
	    $^{1}$College of Information and Computer Sciences, University of Massachusetts \\
	    $^{2}$Department of Computer Science, Emory University\\
	    {\tt mhjang@cs.umass.edu, jinho.choi@emory.edu, allan@cs.umass.edu}
}

\date{}

\begin{document}

\maketitle

\begin{abstract}
Technical documents contain a fair amount of \textit{unnatural language}, such as tables, formulas, pseudo-codes, etc. 
Unnatural language can be an important factor of confusing existing NLP tools.
This paper presents an effective method of distinguishing unnatural language from natural language, and evaluates the impact of unnatural language detection on NLP tasks such as document clustering.
We view this problem as an information extraction task and build a multiclass classification model identifying unnatural language components into four categories.
First, we create a new annotated corpus by collecting slides and papers in various formats, PPT, PDF, and HTML, where unnatural language components are annotated into four categories.
We then explore features available from plain text to build a statistical model that can handle any format as long as it is converted into plain text.
Our experiments show that removing unnatural language components gives an absolute improvement in document clustering up to 15\%.
Our corpus and tool are publicly available.


\end{abstract}

\section{Introduction}

Technical documents typically include meta components such as figures, tables, mathematical formulas, and pseudo-code to effectively communicate complex ideas and results. Let us define the term \emph{unnatural language} as blocks of lines consist of only such components, as opposed to the body text that are natural language.


There are many great NLP tools available as the field has been advanced. However, 
these tools are mostly built for input text that are natural language. As many of our tools for NLP can be badly confused by unnatural language, it is necessary to distinguish unnatural language blocks from natural language blocks, or else unnatural language blocks will cause confusion for natural language processing. Once we salvage natural language blocks from the documents, we can exploit NLP tools much better as they are intended for. This phenomenon is emphasized in technical documents that have higher ratio of unnatural language compared to non-technical documents such as essays and novels.


Document layout analysis aiming to identify document format by classifying blocks into text, figures, tables, and such has been a long-studied problem \cite{gorman,simon}. 
Most previous work have focused on image-based documents, PDF and OCR formats, and used geometric analysis on the pages using the visual cues from its layout. This was a clearly important problem in many applications in NLP and IR.


\begin{figure*}[t]
\centering
\includegraphics[scale=0.3]{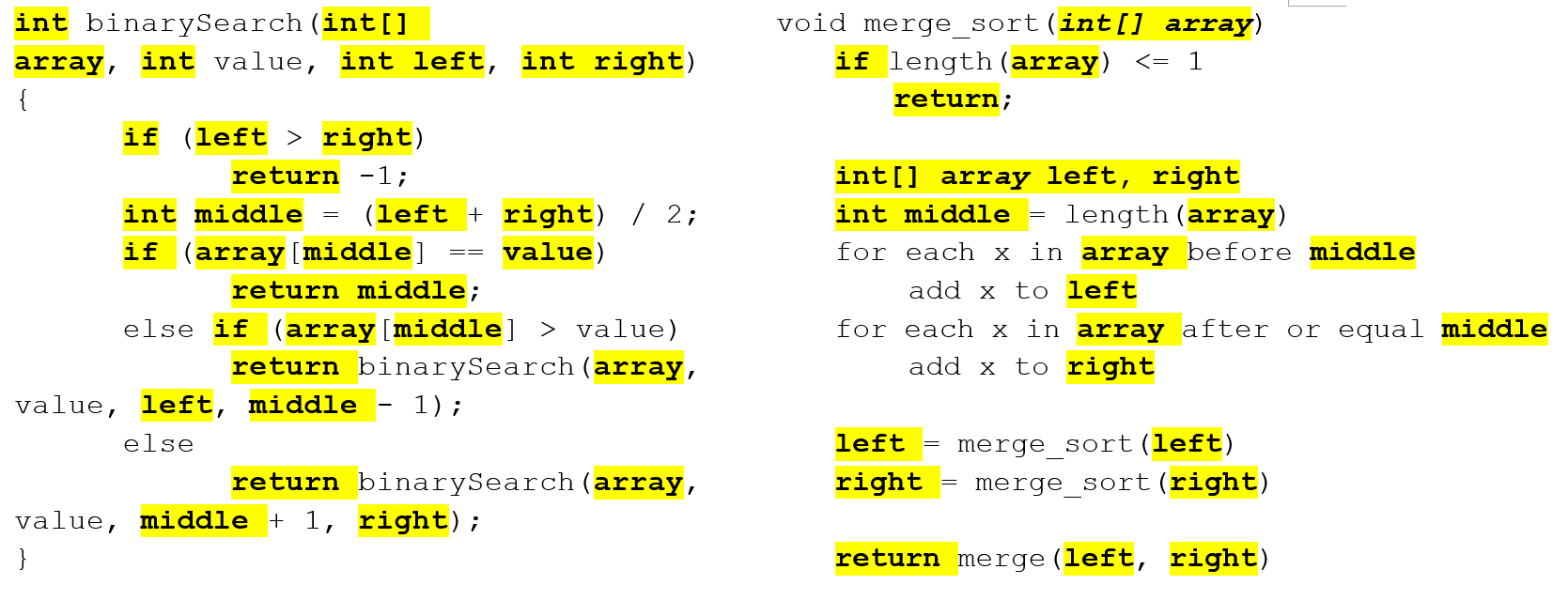}
\caption{An example of how unnatural language confuses NLP tools. The left and right pseudo-codes are very different, but standard NLP similarity functions such as cosine similarity can easily be confused by the terms highlighted in yellow.}
\label{figure:similarcode}
\end{figure*}


This work was particularly motivated while we attempted to cluster teaching documents (e.g., lecture slides and reading materials from courses) in technical topics. We discovered that unnatural language blocks introduced significant noise for clustering, causing spurious matches between documents. For example, code consists of reserved programming keywords and variable names. Two documents can contain two very different code from one another but their cosine similarity is computed high because they share many same terms by programming convention (Figure~\ref{figure:similarcode}). \cite{kohlhase} similarly recognized this problem by explaining main challenges of semantic search for mathematical formula: (1) Mathematical notation is context-dependent; without human's capability to understand the formula from the context, formulas are just \emph{noise}. (2) Identical presentations can stand for multiple distinct mathematical
objects.

This paper proposes a new approach for identifying unnatural language blocks in plain text into four types of categories,(1) \textsc{table} (2) \textsc{code} (3) \textsc{mathematical formula}, and (4) \textsc{miscellaneous (misc.)}. Text are extracted from technical documents in the PDF, PPT, and HTML formats with little to no explicit visual layout information preserved. We focus on technical documents because they have a significant amount of unnatural language blocks (26.3\% and 16\% in our two corpora). Specifically, we focus on documents in slide formats, which have been relatively under-explored. %


We further study how removal of unnatural language improves NLP tasks, document similarity and document clustering.  Our experiments show that clustering on documents with unnatural language removed consistently showed higher accuracy on many of the settings than on original documents, with the maximum improvements up to 15\% and 11\% in two datasets, while it never significantly hurts the original clustering.




\section{Related Work}

\subsection{Table Extraction}
Various efforts have been made for table extraction using semi-supervised learning on the patterns of table layouts within ASCII text documents \cite{ng} web documents \cite{pinto,Lerman01automaticdata,survey} PDF and OCR image documents \cite{pdffigures,tableseer}. Existing techniques exploit the graphical features such as primitive geometry shapes, symbols, and lines to detect table borders. \cite{tablesurvey} introduces and compares the state-of-the-art table extraction techniques from PDF articles. However, there does not appear to be any work that has attempted to process plain text extracted from richer formats, where table layouts are unpreserved.

\subsection{Formula Extraction}
 \newcite{equInpdf} categorized existing approaches for mathematical formulas detection by `character-based' and `layout-based' with respect to key features.  \cite{mer} provides a comprehensive survey of mathematical formula extraction using various layout features available from image-based documents. Since we have no access to layout information, character-based approaches are more relevant to our work. They use features of mathematical symbols, operators, and positions and their character sizes \cite{suzuki,formulafuzzy}. 

\begin{figure*}[t]
\centering
\includegraphics[scale=0.3]{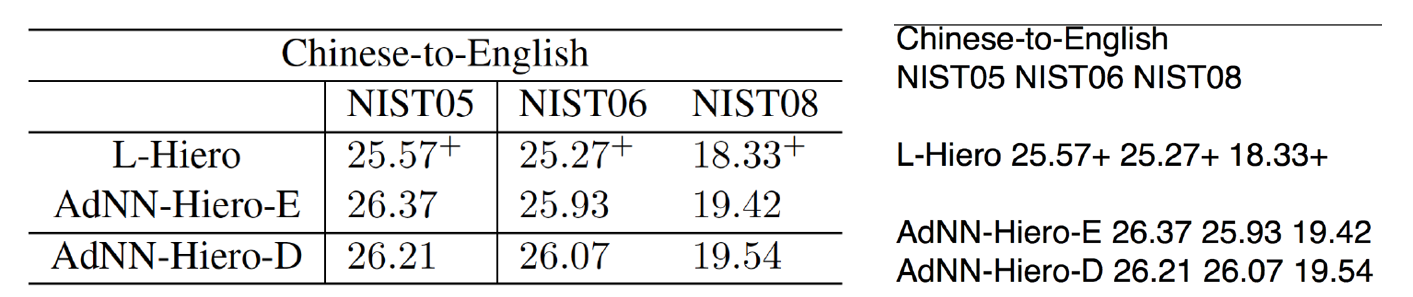}
\caption{A table in PDF document (left) and its text-extracted version (right). Note that it is hard to distinguish the column headings from the extracted text without its layout.}
\label{figure:table-ex}
\end{figure*}

\begin{figure*}[t]
\centering
\includegraphics[scale=0.25]{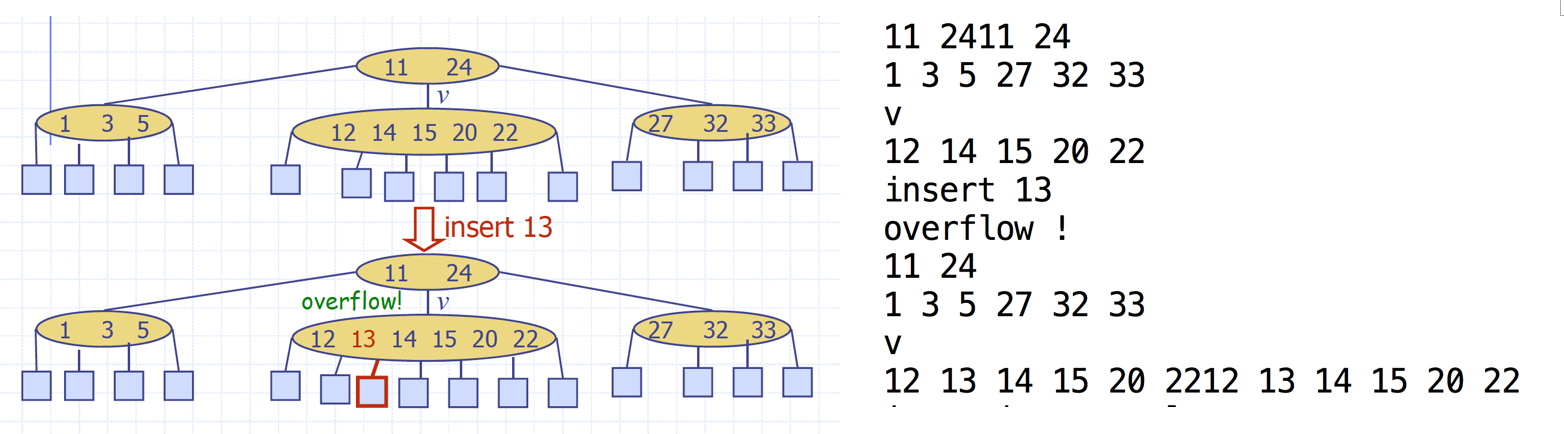}
\caption{Poor text extraction is worse than none. The output from Apache Tika (right) is useless (at best). Experiments will show that document clustering is improved by removing this kind of noise labeled as \textsc{Misc.}}
\label{figure:misc-ex}
\end{figure*}

\subsection{Code Extraction}
\newcite{tuarob} proposed 3 pseudo-code extraction methods: a rule based, a machine learning, and a combined method. Their rule based approach finds the presence of pseudo-code captions using keyword matching. The machine learning approach detects a box surrounding a sparse region and classifies whether the box is pseudo-code or not. They extracted four groups of features: font-style based, context based, content based, and structure based.

\begin{table*}[t]
\centering
\begin{tabular}{@{}l|l|l@{}}
Purpose                                  & Name        & Content                    \\  \hline\hline
\multirow{3}{*}{Classification Training} & T$_{SLIDES}$   & 35 lecture slides (8,514 lines) whose components are annotated                                                        \\
                                         & T$_{ACL}$      & 35 ACL papers (25,686 lines) whose components are annotated    \\
                                         & T$_{COMBINED}$ & Combination of T$_{SLIDES}$ and T$_{ACL}$ \\
                                          \hline
Word Embedding Training                  & T$_{WORD2VEC}$ & \begin{tabular}[c]{@{}l@{}}1,190 lecture slides and 5,863 ACL/EMNLP papers archived \\ over a few years that are used for training word embedding.\end{tabular} \\ \hline
\multirow{2}{*}{Clustering}              & C$_{DSA}$      & 128 lecture slides from `data structure' and `algorithm' classes                                                    \\
                                         & C$_{OS}$       & 300 lecture slides from, `operating system' classes                                                                                                             \\
\end{tabular}
\caption{Datasets used in our paper. All data are available for download at [\url{http://people.cs.umass.edu/~mhjang/publication.html}]}
\label{tab:data-summary}
\end{table*}

\section{Problem Definition}
Input to our task is the plain text extracted from PDF or PPT documents. The goal is to assign a class label to each line in that plain text, identifying it as natural language (regular text) or one of the four types of non-natural language block components, table, code, formula, or miscellaneous text. 
In this work, we focus on these four specific types because our observations lead us to believe they are the most frequently occurring components in PPT lecture slides and PDF articles. Figures are also a very frequent component but we do not consider them because they are commonly pictures or drawings and cannot be easily extracted to text. In this section, we briefly discuss the characteristics of each component and challenges in their identification from the raw text. 

\subsection{Table}
Tables are prevalent in almost every domain of technical documents. Tables are usually conveyed by its two-dimensional layout and its column and/or row headings \cite{tablesurvey}. Figure \ref{figure:table-ex} shows a table in an original PDF document and the same table as it appears the text extracted by Apache Tika\footnote{https://tika.apache.org/}. Tables frequently have multiple cells merged for layout, which makes them particularly difficult to distinguish as table once they are converted to flat text.

\subsection{Mathematical Formula}
Mathematical formulas exist in two ways: isolated formulas on their own lines or as formulas embedded within a line of text. In this work, we treat both types as formula component. Because not all math symbols can be matched to Unicode characters and because the extraction software may not convert them correctly, the extracted text tends to contain more oddly formatted or even completely wrong characters. Superscripts and subscripts are no longer distinguishable and the original visual layout (e.g., math symbols over multiple lines such as $\Pi$ and $\sum$) is lost. 

\subsection{Code}
Articles in Computer Science or related fields often contain pseudo-code or actual program code to illustrate their algorithm. Similar to mathematical formulas, they exist both isolated and embedded, though most code components are isolated code blocks. As in formula component, we treat both types of code blocks as code components. We assume that even indents, one of the strong code visual cues, are not preserved in the extracted text although some extraction tool saves them, not to limit ourselves to the detailed performances of text extraction tools.

\subsection{Miscellaneous Non-text (Misc.)}
In addition to the components mentioned above, there are other types of non-natural language blocks that are left during conversion to text and that may provide spurious sub-topic matches between documents. To allow for those, we denote those components as miscellaneous text. One example of miscellaneous text is the text and caption that are part of the diagrams in slides. Figure \ref{figure:misc-ex} shows an example of miscellaneous text that lost its structure and meaning while being converted to text without the original diagram.

\section{Corpus}
\subsection{Data Collection}
We collected 1,561 lecture slides from various Computer Science and Electrical Engineering courses that are available online, and 5,898 academic papers from several years of ACL/EMNLP archive\footnote{https://aclweb.org/anthology}. We divided the dataset for several purposes; training the classification model, training word embedding model for feature extraction, and clustering for extrinsic evaluation. The details of the dataset we used are summarized in Table ~\ref{tab:data-summary}. We make the data publicly available for download at \url{http://people.cs.umass.edu/~mhjang/publications.html}.

For classification, we constructed three dataset using two different data sources: (1) lecture slides (2) ACL papers, (3) combining both. We chose these two types of data sources because they have different ratios of unnatural language components, hence complementary to each other for the coverage. Table 2 shows the ratio of the four components from each annotated dataset. For example, 1.4\% of lines in $T_{SLIDES}$ are annotated as part of table.

\subsection{Text Extraction}
We extracted plain text from our datasets using an open-source software package, Apache Tika. The package is available for text extraction from various formats including PDF, PPT, and HTML.

\subsection{Annotation}
To train a statistical model, we need ground-truth data. We created annotation guidelines for the 4 types of non-natural language components and annotated 35 lectures slides (7,943 lines) and 35 ACL proceeding papers (25,686 lines). We developed a simple annotation tool to support the task and also to enforce that annotators follow certain rules\footnote{The guidelines and the tool are available at \url{http://people.cs.umass.edu/~mhjang/publication.html}}. We hired four undergraduate annotators who have knowledge of the Computer Science domain for this task.

\setlength{\tabcolsep}{1pt}

\begin{table}[ht]
\label{tab:component_ratio}
\centering
\footnotesize
\begin{tabular}{cccccc}

     & \multicolumn{1}{c}{\textsc{Table}} & \multicolumn{1}{c}{\textsc{Code}} & \multicolumn{1}{c}{\textsc{Formula}} & \multicolumn{1}{c}{\textsc{Misc.}} & 
     \begin{tabular}[c]{@{}l@{}}\scriptsize{All Unnatural}\\ \scriptsize{Categories}\end{tabular}
      \\ \hline
\multicolumn{1}{c}{T$_{SLIDES}$} & 1.4\%  & 14.6\%  & 0.5\%  & 9.8\% & 26.3\% \\ \hline
\multicolumn{1}{c}{T$_{ACL}$}    & 4.0\%  & 0.6\%   & 5.0\%  & 6.4\% & 16\% \\ \hline          
\end{tabular}
\caption{\% of lines by unnatural category. Both datasets have quite a bit of unnatural language (26.3\% for T$_{SLIDES}$ and 16\% for T$_{ACL}$), though T$_{ACL}$ has more \textsc{tables} and \textsc{formulas} and less \textsc{code}.
}
\label{my-label}
\end{table}


\section{Features}
We find line-based prediction has more advantage over token-based prediction because it allows us to observe the syntactic structure of the line, how statistically common the grammar structure is, and how layout patterns compare to neighboring lines. We introduce five sets of features used to train our classifier and discuss each feature's impact on the accuracy.

\subsection{N-gram (N)} Unigrams and bigrams of each line are included as features. 

\subsection{Parsing Features (P)} Unnatural languages are not liklely to form any grammar structure. When we attempt to parse the unnatural language line, the resultant parsing tree would form unusual syntactic structure. To capture this insight, we parsed each line using the dependency parser in ClearNLP \cite{choi:13a} and extracted features such as the set of dependency labels, the ratio of each POS tag, and POS tags of each dependent-head pair from each parse tree. 

\subsection{Table String Layout (T)} Text extracted from tables loses its visual layout as a table but still preserves implicit layout through its string patterns. Tables tend to convey the same type of data along the same column or row. For example, if a column in a table reports numbers, it is more likely to contain numeral tokens in the same location of the lines of the table in parallel. Hence, a block of lines will more likely be a table if they share the same pattern. We encode each line by replacing each token as either S (String) or N (Numeral). We then compute the edit distance among neighboring lines weighted by language modeling probability computed from table corpus (Equation \ref{table-equ}, \ref{table-lm}). 
\begin{multline}
\label{table-equ}
 P_{table}(l_{i}) \propto P_{table}(l_{i} | l_{i-1}) 
\\ = TableLanguageModel(l_{i}) \cdot  \\ editDistance(encode(l_{i}), encode(l_{i-1}))
\end{multline}
\vspace{-4em}

\begin{multline}
\label{table-lm}
  TableLanguageModel(l_i) \\
  = \Pi_{j}^{n}{(P(encode(t_{i,j+1}) | encode(t_{i,j}))} 
\end{multline}
  where $l_i$ refers to a i-th line in a document, $t_{i,j}$ refers to a j-th token in $l_i$.


\subsection{Word Embedding Feature (E)} We train word embeddings using $T_{WORD2VEC}$ using \textsc{word2vec} \cite{embedding}. The training corpus contained 278,719 words. Since we do a line-based prediction, we need a vector that represents the line, not each word. We consider three ways of computing line embedding vector: (1) by averaging the vector of the words, (2) by computing paragraph vector introduced in \cite{le}, (3) using both.

\subsection{Sequential Feature (S)}  The sequential nature of the lines is also an important feature because the component most likely occurs over a block of contiguous lines. We train two models. The first model uses the annotation for the previous line's class. We then train another model using the previous line's predicted label, which is the output of the first model.

\setlength{\tabcolsep}{4pt}

\begin{table*}[ht!]
\centering
\begin{tabular}{c|cccc|cccc}
                    &\multicolumn{4}{c|}{T$_{SLIDES}$}                            & \multicolumn{4}{c}{T$_{ACL}$}                               \\ \hline\hline
                    & \textsc{table} & \textsc{code} & \textsc{formula} & \textsc{misc.} & 
\textsc{table} & \textsc{code} & 
\textsc{formula} & 
\textsc{misc.} \\
W-Random   & 1.69           & 14.62         & 2.82             & 10.57         & 4.15           & 0.62          & 4.44             & 6.08          \\
CLM         & 5.41           & 28.62         & 0.00             & 10.47         & 13.10          & 16.45         & 10.32            & 5.18          \\
Proposed Method & \textbf{67.89}          & \textbf{90.22}         & \textbf{29.09}            & \textbf{89.63}         & \textbf{86.58}          & \textbf{63.70}         & \textbf{80.98}            & \textbf{87.63}        \\
\hline\hline
PC-CB \cite{tuarob}* & N/A & 75.95 & N/A & N/A & N/A & 75.95 & N/A & N/A \\
\end{tabular}
\caption{Single-domain Classification Result in F1-score: Proposed method is much better than baselines for classifying unnatural language. *Note that we borrowed the F1-score reported on their dataset for reference. The number is not directly comparable to other numbers since the datasets are different.}
\label{result:singledomain}
\end{table*}

\section{Classification Experiments}
We used the Liblinear Support Vector Machine (SVM) \cite{svm} classifier for training and ran 5-fold cross-validation for evaluation. To improve the robustness of structured prediction, we adopted a learning to search algorithm known as \textsc{DAgger} to SVM \cite{dagger}.  We first introduce two baselines to compare the accuracy against our statistical model. 

\subsection{Baselines} 
Since no existing work is directly applicable to our scenario, we consider two straightforward baselines.
\begin{itemize} 
\item \textbf{Weighted Random (W-Random)} \\ This assigns the random component class to each line. Instead of uniform random prediction, we made more educated guesses using the ratio of components known from the annotated dataset (Table 2). 

\item \textbf{Component Language Modeling (CLM)} \\ Among the five language models of five component class (the four non-textual components  and text component) generated from the annotations, we predict the component for each line by assigning the component whose language model gives the highest probability to the line. 
\end{itemize}

\subsection{Classification Result} 
We first conducted single-domain classification. Annotations within each dataset, T$_{SLIDES}$ and T$_{ACL}$ are split for training and testing using 5-fold cross validation scheme. Table \ref{result:singledomain} reports F1-score for prediction of the four components in the two dataset using our method as well as baselines.

\begin{table}[h]
\centering
\begin{tabular}{@{}cccc@{}}

\toprule
        & Precision & Recall & F1-score \\ \midrule
Table   & 94.60     & 76.39  & \textbf{84.53}      \\
Code    & 89.56     & 84.01  & \textbf{86.69}      \\
Formula & 85.07     & 79.32  & \textbf{82.10}      \\
Misc.   & 85.59     & 90.24  & \textbf{87.86}      \\ 
Text & 97.76 & 98.79 & 98.27 \\
\bottomrule
\end{tabular}
\caption{Multi-domain classification improves the single-domain classification in Table \ref{result:singledomain}. Identification of categories with particularly low accuracy in each datasets (\textsc{table} and \textsc{formula} in T$_{SLIDES}$ and \textsc{code} in T$_{ACL}$) are improved to be as good as the other categories. }
\label{cross-domain}
\end{table}

Proposed method dramatically increased the prediction accuracies for all of the components against the baselines. CLM baseline showed the highest accuracy on \textsc{code} among the four categories in both datasets. Because pseudo-codes use more controlled vocabulary (e.g., reserved words, common variable names), the language itself becomes distinctive characteristics. We also include the numbers reported by \newcite{tuarob} for comparison. Since their dataset was 258 PDF scholarly articles, T$_{ACL}$ is more comparable dataset than T$_{SLIDES}$, but our training set is much smaller than their dataset. However, their number reported on Table 3 is not directly comparable to other numbers because the numbers are on different datasets.

In T$_{SLIDES}$, the classification F1-score for \textsc{formula} is relatively low as 29.09\% compared to the other components in the same dataset, and also compared to the \textsc{formula} prediction in T$_{ACL}$ (80.98\%). This is due to too small amount of training data (only 0.5\% of \textsc{formula} in T$_{ACL}$), which is overcome in T$_{SLIDES}$ that contain 5\% of \textsc{formula} training data (refer to Table 2).  

In the proposed method, classification of \textsc{code} and \textsc{misc} was significantly improved in T$_{SLIDES}$ (around 90\%), while that of \textsc{table} and \textsc{formula} was improved in T$_{ACL}$ (over 80\%). This shows the complementary nature between two datasets, which suggests that a combined dataset of two, T$_{combined}$, would improve classification performance. Table 4 shows the multi-domain classification result using T$_{combined}$, in which all four categories are identified with F1-score higher than 80\%.

\subsection{Feature Analysis}
We conducted feature analysis to understand the impact of single feature and their combination. We started from single features and incrementally combined them to observe the performance (Figure \ref{figure:figure_analysis}). Features are added in a greedy fashion that a feature that gives the higher accuracy when used alone is added first.

\begin{figure}[t]
\centering
\includegraphics[scale=0.35]{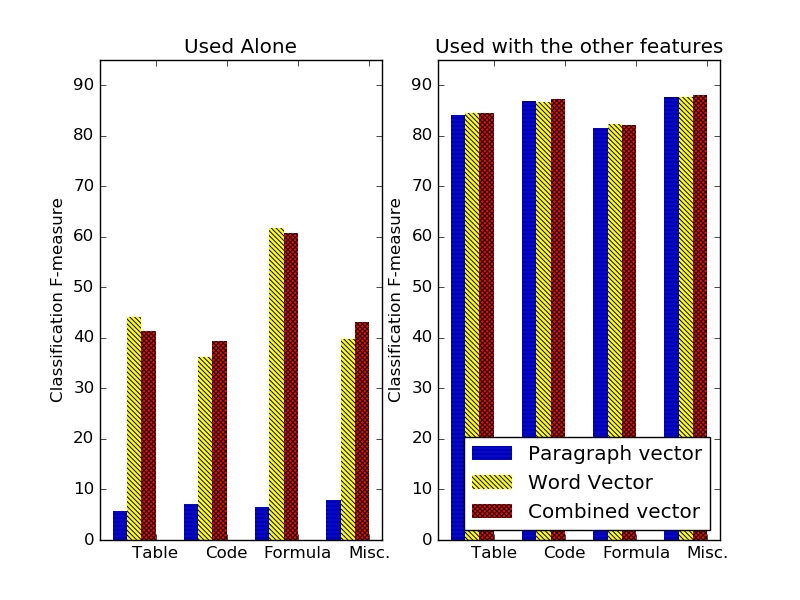}
\vspace{-3mm}
\caption{Three ways of computing sentence embedding vector}
\label{figure:embedding}
\end{figure}

\begin{figure}[h]
\centering
\includegraphics[scale=0.35]{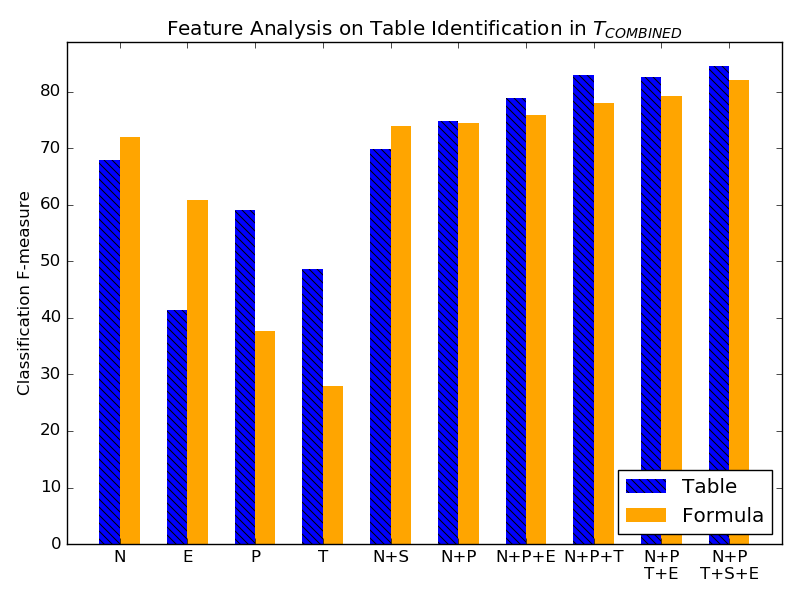}
\vspace{-3mm}
\caption{Feature analysis for \textsc{table} and \textsc{formula}. N: N-gram, E: Embedding, P: Parsing, T: Table String Layout, S: Sequential.  identification in T$_{combined}$}
\label{figure:figure_analysis}
\end{figure}

We first compare the three ways of computing sentence vector features mentioned in Section 5 (Figure \ref{figure:embedding}). When we experiment with only embedding features, averaging word vectors performed 9-12 times better than paragraph vectors. When both features were used, there are some gains in \textsc{Code} and \textsc{Misc.} components and losses in \textsc{Table} and \textsc{Formula}. However, when we experiment with all the other features in addition to embedding features, losses were covered by the other features such that ultimately combined vectors give overall the highest performances.

N-gram (N) features was the most powerful feature with 68\% of F1-score when used alone. The next useful features are parsing feature (P), table layout (T), and embedding features (E) in order for \textsc{table}, while embedding vectors were more effective than parsing feature for \textsc{code} (Figure 5). 






\section{Removal Effects of Unnatural Language on NLP tools} 
We observe how removal of unnatural language from document affects the performance of two NLP tools, document similarity and document clustering. For the set of experiments, we prepared a gold standard clustering for each dataset, C$_{DSA}$ and C$_{OS}$.

\subsection{Document Similarity}
If two documents are similar, they must be topically relevant to each other. A good similarity measure should reflect that; two topically relevant documents should have a high similarity score. To test whether the computed similarity reflects the actual topic relevance better once the unnatural language is removed, we conducted regression analysis.

We converted the gold standard clustering to pair-wise binary relevance. If two documents are in the same ground-truth cluster, they are relevant, and otherwise irrelevant. We then fitted a log-linear model in R for predicting binary relevance from the cosine similarity of document pairs. 

Regression models fitted in R are evaluated using AIC \cite{akaike1974}. The AIC is a measure used as a means for model selection, which measures the relative quality of statistical models learned from the given data. When AIC is smaller, the goodness of fit is better, and the smaller the complexity of the model is, having fewer parameters to represent the model. Table \ref{tab:regression} shows that AIC was reduced by 53 and 118 respectively on the models trained with documents whose unnatural language blocks are removed, compared to the original documents. Since AIC does not provide a test for a model, AIC does not suggest anything about the quality of the model in an absolute sense, but relative quality. From this result, we can conclude that cosine similarity can fit a better model that predicts documents' topic relevance with significance after unnatural language blocks have been removed.

\begin{table}[h]
\centering
\small
\begin{tabular}{c|c|c|c}

    & AIC(D$_{original}$) & AIC(D$_{removed}$) & Improvement \\ \hline
C$_{DSA}$ &   -40975   &     -41028    & -53             \\
C$_{OS}$  &    -61404  & -61522    & -118           
\end{tabular}
\caption{
The statistical model is trained better with documents whose unnatural language categories are removed (D$_removed$) than the model with the original documents (D$_original$) in both datasets.  \emph{Smaller} AIC scores imply \emph{better} models.}
\label{tab:regression}
\end{table}

\begin{figure}[t]
\centering
\includegraphics[scale=0.25]{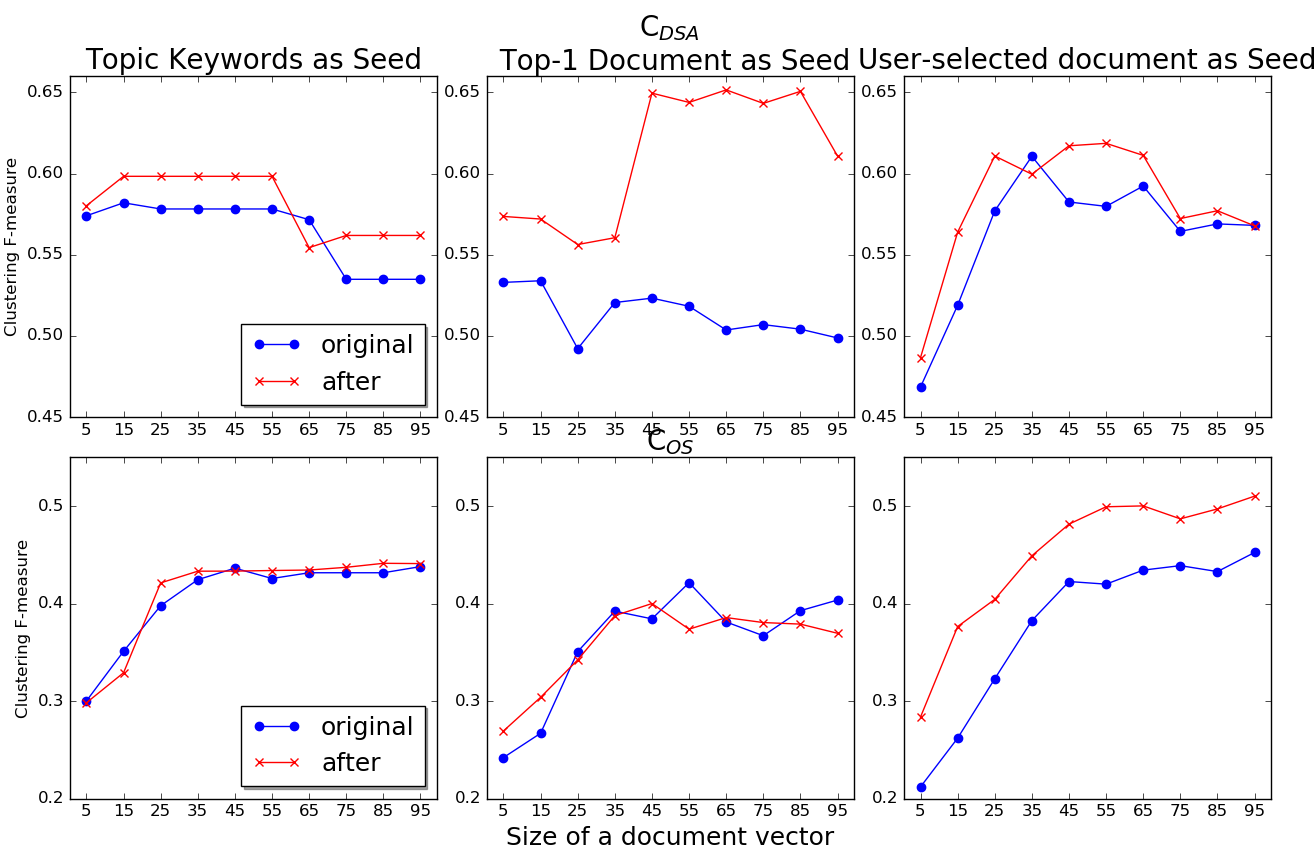}
\caption{Clustering result on two datasets, C$_{DSA}$ (top) and C$_{OS}$ (bottom). X axis referes to the the size of document vector K, which controls the top-K TF-IDF terms included from documents. Y axis: Clustering F1-score. }
\label{figure:clustering}
\end{figure}

  
  \vspace{-2em}

\subsection{Document Clustering} 
Comparing general clustering performance on two document sets is tricky because clustering performance varies by many factors, e.g., clustering algorithm, similarity function, document representation, and parameters. To make a safe claim that clustering quality of one set of documents is better than of the other, clustering on one set should consistently outperform the other under many different settings. To validate this, we perform clustering experiments with multiple settings such as varying document vector size and and different initialization schemes.

In this experiment, we consider seeded K-means clustering algorithm \cite{kmeans} for teaching documents. In our application scenario, users initially submit a topic list (e.g., syllabus) of the course. Then lecture slides are grouped into the given topic cluster. Depending on users' interaction level, we consider a semi-interactive scenario where users only provide a topic list, and a fully-interactive setting where users not only provide a topic list but also provide an answer document for each topic cluster, more specifying the intended topic.

\begin{algorithm}[ht]
\label{seeded_kmeans}
 \KwData{Set of document vectors D = \{$d_1, ... d_n$\}, $d_i \in R^T$, set of seed vectors S = \{$s_1, ...s_k$\}, user-provided topic keywords vector T = $\{t_1,...t_k\}$}  \KwResult{Disjoint K partitioning of D into C$_{l=1}^{k}$}
 
 {Seed Initialization:}\\
 \phantom{haa}(1) Topic-keywords seeding:    $s_i = t_i$,  \\
 \phantom{haa}(2) Top-1 document seeding: \\  
 \phantom{haaa}$s_i = d_j$  $argmax_j(\textsc{cosineSimilarity}(t_i, d_j))$ \\
 \phantom{haa}(3) User-selected document seeding: \\  
 \phantom{haaa}$s_i = $\textsc{DocselectedByUser}($t_i)$  \\
 \While{convergence}{
  K-means clustering document selection process
 }
 \caption{Seeded K-means with User Interaction}
\end{algorithm}

In a semi-interactive setting, topic keywords are sparse seed as they usually consist of two or three words. Therefore, we expand the topic keywords by finding the top-1 document retrieved from the keywords and use it as seed. For experiments, we simulate the fully-interactive setting; instead of having an actual user to pick an answer document, we use an answer document randomly chosen from a gold cluster. The seeded K-means clustering algorithm with three interactive seeding schemes is described in Algorithm 1.


We can consider a simulated setting more realistic when the selected document is suggested to the user as the top or a near-top choice. In our dataset, 60\% of the selected documents were ranked in top 10 
in C$_{DSA}$, and 13\% of the selected documents were ranked in top 10 in C$_{OS}$, 
which implies that the simulated setting in C$_{DSA}$ was more realistic than in C$_{DSA}$. For top-1 document seeding, 64\% and 78\% of document seeds matched with the gold standard in C$_{DSA}$ and C$_{OS}$, respectively. 

Figure \ref{figure:clustering} shows the clustering result of original documents (D$_{original}$) and documents whose unnatural language blocks removed (D$_{removed}$), with three different seeding schemes over two lecture slides datasets. In C$_{DSA}$, D$_{removed}$ consistently outperformed with all three seeding schemes. The clustering performed the best with D$_{removed}$ when top-1 document was used as seed. Overall, in C$_{DSA}$, clustering was improved 94\% of the times with the maximum absolute gain of 14.7\% and the average absolute gain of 4.6\%. The average absolute loss was 0.8\% when 6\% of the times it hurt. In C$_{OS}$, clustering was improved 73\% of the times with the maximum absolute gain of 11.4\% and the average absolute gain of 3.9\%. The average absolute loss was 1.7\%. Our results suggest that removal of unnatural language blocks can significantly improve clustering most of the times with bigger gain than occasional losses.

\vspace{-0.5em}
\section{Conclusion}

In this paper, we argued that unnatural language should be distinguished from natural language in technical documents for NLP tools to work effectively. We presented an approach to the identification of four types of unnatural language blocks from plain text, which is not dependent a document format. Proposed method extracts five sets of line-based textual features, and showed above 82\% F1-score for the four categories of unnatural language.  
We showed how existing NLP tools can work better on documents if we remove unnatural language from documents. Specifically, we demonstrated removing unnatural language improved document clustering in many settings by up to 15\% and 11\% at best, while not significantly hurting the original clustering in any setting. 



\section*{Acknowledgments}
This work was supported in part by the Center for Intelligent Information Retrieval, in part by NSF grant \#IIS-0910884, and in part by NSF grant \#IIS-1217281. Any opinions, findings and conclusions or recommendations expressed in this material are those of the authors and do not necessarily reflect those of the sponsor. The authors thank Kenneth W. Church for providing valuable comments and advice. 

\bibliography{naaclhlt2016}
\bibliographystyle{naaclhlt2016}

\end{document}